# Chameleon: A Color-Adaptive Web Browser for Mobile OLED Displays

Mian Dong and Lin Zhong
Department of Electrical and Computer Engineering, Rice University

**Abstract**

Displays based on organic light-emitting diode (OLED) technology are appearing on many mobile devices. Unlike liquid crystal displays (LCD), OLED displays consume dramatically different power for showing different colors. In particular, OLED displays are inefficient for showing bright colors. This has made them undesirable for mobile devices because much of the web content is of bright colors.

To tackle this problem, we present the motivational studies, design, and realization of *Chameleon*, a color adaptive web browser that renders web pages with power-optimized color schemes under user-supplied constraints. Driven by the findings from our motivational studies, Chameleon provides end users with important options, offloads tasks that are not absolutely needed in real-time, and accomplishes real-time tasks by carefully enhancing the codebase of a browser engine. According to measurements with OLED smartphones, Chameleon is able to reduce average system power consumption for web browsing by 41% and reduce display power consumption by 64% without introducing any noticeable delay.

## 1. Introduction

Displays are known to be among the largest power-consuming components on a modern mobile device [1-3]. OLED displays are appearing on an increasing number of mobile devices, e.g., Google Nexus One, Nokia N85, and Samsung Galaxy S. Unlike LCDs where the backlight dominates the power consumption, an OLED display does not require backlight because its pixels are *emissive*. Each pixel consists of several OLEDs of different colors (commonly red, green and blue), which have very different luminance efficiencies. As a result, the color of an OLED pixel directly impacts its power consumption. While OLED displays consume close to zero power when presenting a black screen, they are much less efficient than LCDs in presenting certain colors, in particular white. For example, when displaying a white screen of the same luminance, the OLED display on Nexus One consumes more than twice of that by the LCD on iPhone 3GS [4]. Because the display content usually has a white or bright background, OLED displays are considered less efficient than LCDs overall. For example, Samsung reportedly dropped OLED for its Galaxy Tablet due to the same concern [5].

Our goal is to make web browsing more energy-efficient on mobile devices with OLED displays. Not only is web browsing among the most used smartphone applications according to recent studies [6, 7], but also most of today's web content is white (80% according to [8]). For example, about 60% of the average power consumption by Nexus One of browsing CNN mobile is contributed by the OLED display according to our measurement. Our algorithmic foundation is prior work by us [9, 10] and others [11] that has demonstrated the potential of great efficient improvement in OLED displays by changing the color of display content, or *color transformation*. In order to effectively apply color transformation to web pages for energy efficiency we must answer the following questions.

First, *will web content providers solve the problem by providing web pages in energy-efficient color schemes?* Some web sites allow the end users to customize the color schemes; and many provide a mobile version. However, our study of web usage by smartphone users revealed that close to 50% of web pages visited by mobile users are not optimized for mobile devices at all [12]. Therefore, while it would be ideal if each web site provides a version of content optimized for OLED displays, it is unlikely to happen at least in the near future. Moreover, our measurement of OLED smartphones showed that different OLED displays may have different color-power characteristics. There is no single color scheme that is optimal for all OLED displays.

Second, *will the problem be solved by configuring a web browser with the most energy-efficient color scheme?* Some web browsers already allow users to customize their color schemes. A user can simply choose the most energy-efficient scheme. This solution is, however, highly limited because the customized color scheme only affects the web browser itself, not the content rendered by it. Some web browsers allow a user-defined color style to set the color for texts, backgrounds, and links in the web content. Unfortunately, our study shows that an average web page visited by smartphone users employs about four colors for texts and three for backgrounds. If a single user-defined color style is applied, many web pages will be rendered unusable.

Finally, *is it feasible to implement color transformation at the mobile client?* Color transformation requires collecting statistics of color usage by pixels in real-time, transforming each color, and applying the new color to web page rendering in real-time. Because both the number of pixels (~$10^5$) and the number of colors (~$10^7$) are large, a straightforward application of color transformation will be extremely compute-intensive [9, 10] and, therefore, defeat the purpose of energy conservation.

By realizing Chameleon, we answer the last question affirmatively. The key to Chameleon's success lies in a suite of system techniques that leverage the findings from our motivational studies. Chameleon applies color transformation to web contents with the following features. It applies consistent color transformations to web pages from the same web site and applies different perceptual constraints in color transformations for web content of different fidelity requirements. Chameleon constructs the power model of an OLED display without external assistance in order to achieve device-specific optimal color transformation. It allows end users to set their color and fidelity preferences in color transformation. Finally, Chameleon only performs the absolutely necessary tasks in real-time and aggressively optimizes the real-time tasks for efficiency. Our evaluation shows that Chameleon can reduce the average system power consumption during web browsing by 41% without introducing any user noticeable delay on OLED smartphones.

In designing and realizing Chameleon, we make the following contributions:

- Three motivational studies that lead to the design requirements of Chameleon (Section 3).
- The design of Chameleon that extensively leverages the findings from the motivational studies in order to meet the design requirements (Section 4).
- Efficient realizations of Chameleon based open-source web browsers, including Android Webkit and Fennec (Section 5).

A very early prototype of Chameleon was demonstrated at HotMobile'10 [13].

## 2. Background and Related Work

We next provide background and discuss related work.

### 2.1 Color Spaces

A color sensation by human can be described with three parameters because the human retina has three types of cone cells that are most sensitive to light of short, middle, and long wavelengths, respectively. A color space is a method for describing color with three parameters. Most used color spaces include linear RGB, sRGB and CIELAB.

The linear RGB and sRGB spaces are designed to represent physical measures of light. In the linear RGB color space, a color is specified by $(R, G, B)$, the intensity levels of the primary colors: red, green and blue, which will create the same color sensation when combined. In the sRGB (standard RGB) color space, a color is also specified by $(R, G, B)$, but the intensity levels are transformed by a power-law compression, or gamma correction, to compensate the non-linearity introduced by conventional CRT displays. Although CRT displays are no longer common nowadays, the sRGB color space is still widely used in electronic devices and computer displays. By default, the RGB values used in almost all mainstream operating systems and applications are in the sRGB color space.

The CIELAB color space is designed to mimic the human vision. In the CIELAB color space, a color is specified by $(L^*, a^*, b^*)$, where $L^*$ represents the *lightness*, human subjective brightness perception of a color, while $a^*$ and $b^*$ determine the *chromaticity*, the quality of a color. The lightness of a color can be calculated from its relative luminance to a standard white point [15]. The CIELAB color space is so designed that uniform changes of $L^*a^*b^*$ values aim to correspond to uniform changes in human perception of color. As a result, the Euclidean distance in the CIELAB color space is usually used to measure color difference perceived by human [15].

### 2.2 OLED Display and Color Transformation

Organic light-emitting diode or OLED [16, 17] technology promises much better dynamic color, contrast and a much thinner, lighter panel than conventional LCDs. Unlike LCDs, an OLED display does not require external lighting because its pixels are emissive. Each pixel of an OLED display consists of several colorful OLEDs, usually red, green and blue, respectively. Because the red, green, and blue components of a pixel have different luminance efficacies, the color of a pixel directly impacts its power consumption. In contrast, illumination of backlight, not color, determines the power consumption by an LCD.

#### 2.2.1 Color-Power Model

Knowing the color of all pixels of an OLED display region, one can calculate the power consumption contributed by the region readily. Assume the display region of interest contains $N$ colors, $x_1, x_2, \ldots, x_N$, each specified by a three-element column vector, either in the linear RGB or CIELAB space, i.e., $x_i = (R_i, G_i, B_i)^T = (L_i^*, a_i^*, b_i^*)^T$. Using results reported in our prior work [18], one can count the number of pixels for each color, i.e., $n_i$ for color $x_i$, and calculate the power consumption of the region as

$$P = \sum_{i=1}^{N} n_i \cdot P_{pixel}(x_i),$$

where $P_{pixel}(x)$ is the pixel power model.

#### 2.2.2 Display Darkening

Current OLED smartphones, e.g., Nexus One and Galaxy S, provide a display darkening mechanism by which users can change the brightness of the whole screen uniformly. The visual effect is similar to that of backlight dimming on an LCD. Such a mechanism is implemented in driver circuit of the OLED panel and is uniformly applied to the whole display without digitally changing the color of a pixel. Display darkening is less effective in power reduction than color transformation is because display darkening only changes lightness of colors while color transformation changes both lightness and chromaticity. As will be shown in Section 3.3, given the same perception constraints, color transformation is able to reduce display power consumption by three times more than display darkening is.

*2.2.3 Color Transformation*

The display darkening described above can be considered as a special case of color transformation [9]. Color transformation considers both the lightness and chromaticity of a color and transforms colors one by one, instead of uniformly. Our early results presented in [9, 10] show that color transformation can significantly reduce the display power consumption without sacrificing user satisfaction.

The objective of color transformation is to find a *color map*, or $N$ transformed colors, $x'_1, x'_2, \ldots, x'_N$, to substitute the original $N$ colors, $x_1, x_2, \ldots, x_N$, such that the display power consumption is minimized, while meeting a perception constraint. Generally speaking, there are two types of perception constraints, constraint in fidelity and constraint in usability.

When fidelity matters for the screen region, e.g., in the case of photos, the distortion introduced by the color transformation can be guaranteed to be below a user-specified threshold $\delta$. The distortion can be measured as the total pixel by pixel color difference between the original and transformed screens [19], i.e.,

$$\sum_{i=1}^{N} n_i \|x_i - x'_i\|_2 \leq \delta.$$

When usability instead of fidelity matters, e.g., in the case of GUI objects, the color difference between any two colors after transformation can be guaranteed to be close to the color difference of the two original colors in order to preserve contrast and features. For example,

$$\forall i,j \in \{1,2,\ldots N\}, \|x'_i - x'_j\|_2 = \lambda \|x_i - x_j\|_2$$

where $\lambda$ is a user-supplied parameter. Please see our prior work [9, 10] for a complete treatment.

Chameleon supports two forms of color transformations that are subject to user choices. In an *arbitrary transformation*, each color can be transformed into any color. Arbitrary transformation is better where colors are independent from each other in the original screen, e.g., solid color areas in a GUI. Moreover, an arbitrary transformation can potentially achieve the maximum power reduction, given the perceptual constraint. In a *linear transformation*, the same linear function is identified and applied for all colors in the same screen. A linear transformation maps all colors with the same linear function. Linear transformations are better where colors are dependent on each other, e.g., color gradients. The linearity will preserve the relative positions of colors in the color space such that the gradients will also be preserved.

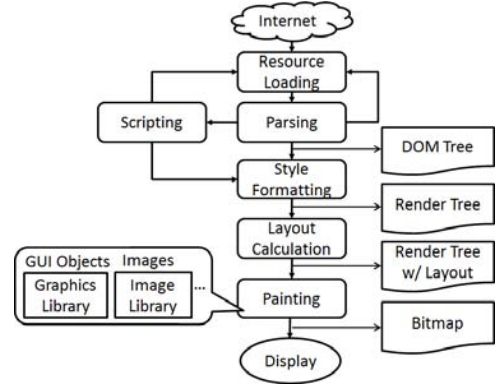

**Figure 1: A simplified workflow of a web browser**

### 2.3 Web Browser

Modern web browsers render a web page through a complicated process teemed with concurrency and data dependency. Figure 1 shows the workflow of a web browser. To open a web page, the browser first loads the main HTML document and parses it. When parsing the HTML document, more resources such as other HTML documents, Cascading Style Sheets (CSS), JavaScripts, and images, may be discovered and then loaded. These two iterative stages are *Resource Loading* and *Parsing* in Figure 1. In *Parsing*, the browser manipulates objects specified by HTML tags in the web page using a programming interface called Document Object Model (DOM). These objects are therefore known as *DOM elements*, which are stored in a data structure called the *DOM tree*.

In *Style Formatting*, the browser processes CSS and JavaScripts to obtain the *style* information, e.g., color and size, of each DOM element and constructs a *render tree*, the visual representation of the document. Each node in the render tree is generated from a DOM element to represent a rectangular area on the screen showing the element. The style information of each DOM element is stored as *properties* of the corresponding node in the render tree.

Then, in *Layout Calculation*, the browser computes the layout and updates the position property of each node in the render tree based on the order of these nodes. Chameleon utilizes position and size properties of image nodes to identify which pixels in the current screen belong to images. This is discussed in Section 4.4.

Finally, in *Painting*, the browser calls a series of paint functions to draw the nodes of the render tree with layout onto a bitmap in the framebuffer to represent the screen and each paint function covers a group of adjacent nodes on the screen. Chameleon catches these paint functions to identify which regions of the current screen are updated. Nodes of different types are painted using different libraries. Images are painted using various image libraries depending on the image format. GUI objects, e.g., links, forms, and tables, etc, are painted using a graphics library that handles basic geometry elements such as points, lines and rectangles. Chameleon realizes color transformation by modifying the interfaces of images and graphics libraries.

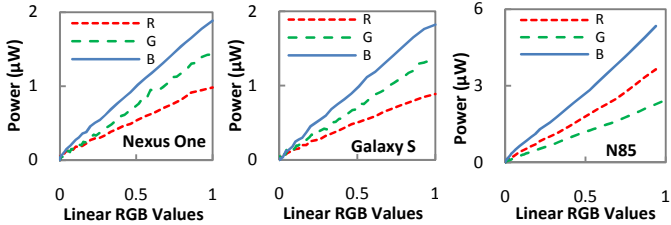

Figure 2: Power models of three OLED displays. OLED display power model is a linear function of linear RGB values. Different OLED displays have different power models

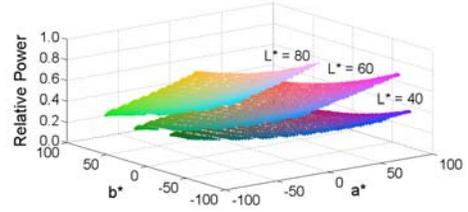

Figure 3: Power vs. CIELAB. Given the lightness, or L*, power consumption can differ as much as 5X between two colors with different chromaticity

### 2.4 Related Work

Chameleon is motivated by existing work in display power management and leverages its algorithmic solutions. HP Labs pioneered energy reduction for OLED-based mobile displays [11] by darkening the display regions that is outside the focal area. User studies [20] showed that this technique has high user acceptance for displaying notifications and menus but low user acceptance for tasks like reading messages and books because it is hard to determine the user's focal area in these tasks. Web browsing is, unfortunately, similar to the latter. In contrast, Chameleon does not need to know the focal area and is more effective in conserving power thanks to color transformation.

The IBM Linux Watch is one of the earliest users of OLED-based displays. Its designers sought to use low-power colors for more pixels and designed GUI objects, such as fonts and icons to minimize the need for high-power colors [21]. The work, however, only studied the GUIs with two colors, i.e., background and foreground, without addressing colorful designs. There is a large body of work on energy optimization of conventional LCD systems. It reduces external lighting and compensates the change by transforming the displayed content [19, 22, 23]. Most of these techniques can be applied to OLED-based displays by scaling the luminance level of OLED pixels as discussed in Section 2.2. Chuang *et al.* studied energy-aware color transformation for LCDs [24], targeted for data visualization applications. Our prior work, reported in [9, 10], was the first work that applied color transformation to GUIs on OLED displays under usability perceptual constraints. Both pieces of work provide algorithms that can be used in Chameleon to transform colors in web pages.

There exists a lot of research effort that adapts web pages for mobile displays either manually [25] or automatically [26]. Its goal to better fit web content into the small mobile display is very different from ours in conserving energy. Moreover, its solutions usually involve the modification of layout, instead of color.

## 3. Motivational Studies

We next report three studies that directly motivated the design of Chameleon: the OLED displays, web usage by smartphone users, and user preferences in web page color transformation.

### 3.1 OLED Display Power

Using the procedures described in [18], we build pixel level power models of three OLED smartphones, i.e., Nexus One, Galaxy S, and Nokia N85. The results are shown in Figure 2. We make the following observations as related to the design of Chameleon.

*First, OLED display power model is a linear function of linear RGB intensity levels.* The linear regression fitting statistics $R^2$ of all the three devices are over 0.95. The reason is that the power consumption of an OLED is a linear function of the current passing through it. And the current passing through an OLED is also a linear function of its luminance represented by linear RGB intensity levels [27]. The linearity simplifies the construction of OLED power model as is necessary in Chameleon.

*Second, different displays have different power characteristics.* Particularly, relative power consumption by red and green colors varies significantly from device to device. In N85, red is more power efficient than green; in Nexus One, the opposite is true, as shown in Figure 2. This means that the most energy-efficient color scheme on N85 may be not most energy-efficient on Nexus One. This observation motivates a device-specific color transformation that employs a device-specific OLED power model.

*Finally, chromaticity makes a big difference even the lightness is identical.* Figure 3 presents the power model of the OLED display of Nexus One in CIELAB color space, in which power consumption of each color is normalized by that of white. Each curved surface in the figure represents of the colors of the same lightness ($L^*$). Unsurprisingly, the figure shows that the power consumption of a color will increase when lightness increases given the same chromaticity ($a^*$ and $b^*$). This indicates that one should use darker colors to reduce power consumption of an OLED display, which has been widely known and practiced already. More importantly, however, Figure 3 also shows that the power consumption difference can be as high as 5X between two colors of the same lightness, or on the same curved surface. This finding indicates that changing chromaticity is another effective way to reduce OLED display power and color transformation will be more effective than display darkening.

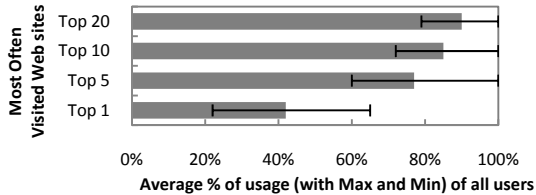

**Figure 4: The most visited web sites accounts for a high percentage of web usage**

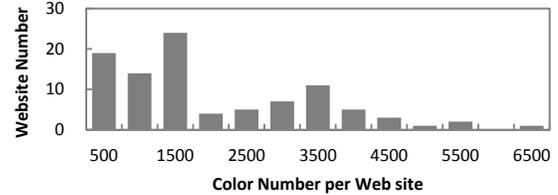

**Figure 5: Color number distribution. A web site uses about 1500 colors on average and 6500 at most**

## 3.2 Web Usage by Smartphone Users

By studying web browser traces from 25 iPhone 3GS users over three months [6] (called *LiveLab traces* below), we make the following observations as related to the design of Chameleon.

### 3.2.1 Browsing Behavior

*First, mobile users still visit web pages that are not optimized for mobile devices.* While a web site can potentially provide an OLED-friendly version, e.g., with dark background, our data show that approximately 50% of web pages visited by mobile users are not optimized for mobile devices at all [12]. Therefore, one cannot count on every web site to provide an OLED-friendly version to reduce display power. This directly motivates the necessity of client-based color transformation.

*Second, a small number of web sites account for most web usage..* We find that the 20 most often visited web sites of each user contribute to 80%-100% (90% on average) of the web usage by the same user, as shown in Figure 4. Therefore, it is reasonable to maintain a color transformation scheme for each of the 20 web sites and to use a universal transformation scheme for, or simply not transform, the other web sites. This is the key rationale behind our design decision to maintain color consistency per web site in Chameleon (Section 4.1).

### 3.2.2 Web Content

We further analyzed the web pages visited by the 25 iPhone 3GS users with the following findings.

First, *65% of the pixels in the web pages visited by the 25 users over three months are white.* This is different from but close to the claim made by [8] that white color takes as high as 80% of web content. As OLED displays are power-hungry for white, they can be less energy-efficient overall than LCDs. Color transformation is therefore very important to improve the energy efficiency of mobile devices with an OLED display.

Second, *an average web page from the LiveLab traces includes very rich styles*, about *four* colors for texts and *three* for backgrounds. On the contrary, a user defined Cascading Style Sheet (CSS), or browser color profile only defines one color for all texts and one color for all backgrounds. As a result, using a user defined CSS to format an entire web page will significantly reduce color numbers of the web page and compromise aesthetics or even impact the web usability, as exemplified by Figure 6 (b).

Third, *the number of colors used by a web site is very small ($\sim 10^3$) compared to the number of all colors ($\sim 10^7$) supported by a modern mobile device.* The number of colors has a significant implication on the computational cost of color transformation not only because each color has to be transformed but also because the occurrences of each color must be counted for an optimal transformation. The number of colors is as many as $2^{24}$ or 16,777,216, in a modern web browser in which each of RGB components is represented using eight bits. While this number is prohibitively high, we find that web sites accessed by LiveLab users only have about 1500 colors on average, with the maximum being 6500, as shown in Figure 5. This observation is key to Chameleon's feasibility and design of color contribution collection.

Finally, *a modern web page contains visual objects of different fidelity requirements in color transformation.* Videos and many images require color fidelity. That is, their color cannot be modified arbitrarily. An analysis of the LiveLab traces shows that such fidelity-bound objects are abundant in web pages accessed by mobile devices. For example, images account for about 15% of the pixels for an average web page in the trace. As a result, blindly changing the color of a pixel without considering the fidelity requirement of a visual object, e.g., inverting all pixels in the page (Figure 6 (c)), can be unacceptable. In contrast, GUI objects only require usability. Here, *GUI objects* are all possible objects in a web page except images and videos, including backgrounds, texts, and forms etc, which cover approximately 85% of the pixel of a web page based on our LiveLab traces analysis. Their colors can usually be modified arbitrarily as long as the transformed colors are still distinguishable and aesthetically acceptable.

Similarly, *images on a web page may have different fidelity requirements too. Foreground images* are images specified by HTML IMG tags. For most foreground im-ages, such as photos, fidelity matters. Significant color changes will render them useless as shown in Figure 6 (c). However, logo images, a common, special kind of foreground images, usually have the same background color as the adjacent GUI objects by design. As a result, if a logo image is not transformed along with the GUI objects, the background color of the logo image will clash with the inverted background color, as shown in Figure 6 (e). Therefore, Chameleon allows the user to choose if it is important to keep the fidelity of logo images. If not, Chameleon will treat logos along with the GUI objects in color transformation. *Background images* are images specified by CSS BACKGROUND-IMG property to serve as the background of a CSS box. In the LiveLab traces, 23% of web pages contain background images. Because background images are usually designed with the same color pattern as the adjacent GUI objects, color transformation without background images may make the background images undistinguishable from adjacent GUI objects with transformed colors. As shown in Figure 6 (e), the background images of the two buttons are undistinguishable from inverted texts. Therefore, Chameleon treats background images along with the GUI objects in color transformation.

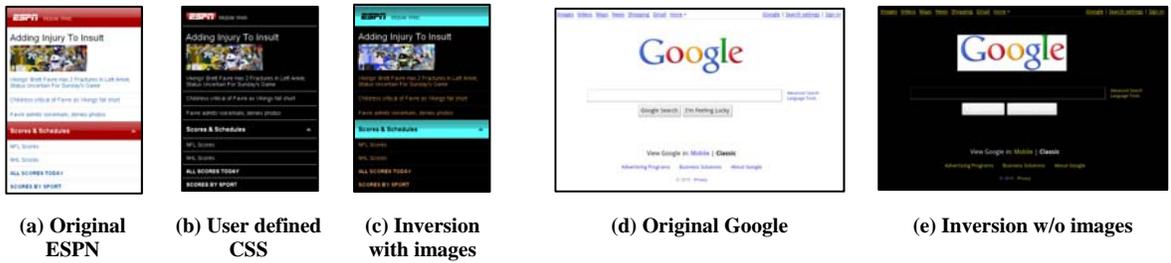

| (a) Original ESPN | (b) User defined CSS | (c) Inversion with images | (d) Original Google | (e) Inversion w/o images |

**Figure 6:** The original ESPN home page (a) and its three color transformations (b-c). The use of a user defined CSS (b) loses color information of different CSS boxes. *Inversion* (c) makes images unusable. In Inversion w/o images (e), background images are undistinguishable from inverted texts in two buttons and foreground logo image clashes with the inverted background color

### 3.3  User Preference of Color Transformation

While color transformation can potentially reduce the power consumption by an OLED display, it must be performed with user acceptance considered. We employ a series of user studies to investigate how users accept color transformation of web pages. We recruited 20 unpaid volunteers to use Nexus One to review a series of web pages with typical office lighting. The participants were asked to score each web page by 1 to 5 with 1 being the least acceptable.

The web pages used in the experiment include four pages from the five top mobile web sites, i.e., CNN, Facebook, Google, Weather, and ESPN. For each web page, we present the original and four color-transformed versions, as shown in Figure 7. Thus, there are 100 pages in total. The color transformations include:

- *Dark*: Lightness of all the colors is uniformly reduced, i.e., $R' = \lambda R; G' = \lambda G; B' = \lambda B$, in which $\lambda \in [0, 1]$. This is similar to what a user would experience with modern smartphones with LCDs and OLED displays.

- *Green*: Lightness of each of the RGB channel is reduced separately, i.e., $R' = \lambda_R R; G' = \lambda_G G; B' = \lambda_B B$, in which $\lambda_R, \lambda_G, \lambda_B \in [0, 1]$. Green channel is reduced least because green is the most efficient color.

- *Inversion*: All the colors are inverted by replacing each of the RGB components with its complement multiplying a scalar $\lambda$, i.e., $R' = \lambda(1 - R); G' = \lambda(1 - G); B' = \lambda(1 - B)$, in which $\lambda \in [0, 1]$. The rationale of this transformation is that most pixels from web pages are white, according to Section 3.3.

- *Arbitrary*: Lightness and chrome of each color are changed to minimize display power consumption [9].

The transformations are only applied to GUI objects, background images and non-photo foreground images. Each of these algorithms includes one or more controllable parameters, e.g., lightness reduction ratio $\lambda$ in *Dark*. Such parameters affect both perception constraints and power consumption of transformed web pages. For a fair comparison, we adjust the parameters in each transformation so that the same usability constraint discussed in Section 2.2.3 is used in all four algorithms.

*Dark*, *Green*, and *Inversion* are linear transformations as discussed in Section 2.2.3. As a result, *Arbitrary* promises the biggest power reduction. Not, surprisingly, the power reduction by *Dark*, *Green*, *Inversion*, and *Arbitrary* is 25%, 34%, 66%, and 72%, respectively, under the same perceptual constraint.

We have the following two findings by analyzing the scores by the participant. *First, different users prefer different transformations for a web site.* For each color transformation of a web page, we count the number of users who gave the highest score out of the four transformations. Thus, we have four numbers of user "votes" for each web page. Figure 8 (left) shows the four numbers for the homepage of each web site used in our study. As shown in the figure, given a web site, each transformation gets some votes. The numbers of votes for all four transformations are actually not dramatically different for most web sites. Therefore, it is important to give end users options in selecting the color transformation algorithm, instead of providing one for all users.

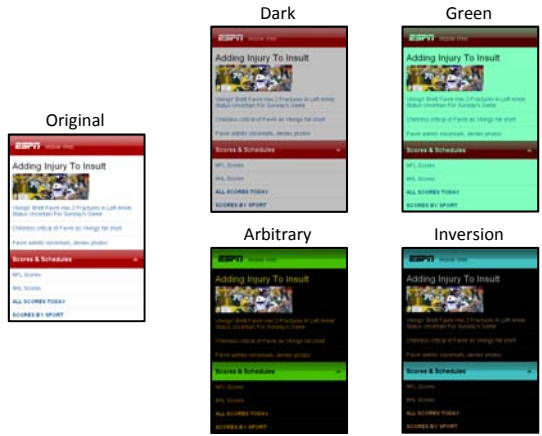
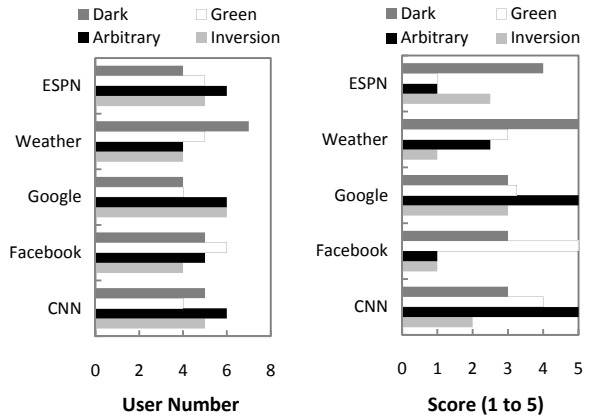

Figure 7: Color schemes used in user study are transformed using different algorithms

Figure 8: User study results. User favorite algorithm (left); average score of each web site by user1 (right)

*Second, even the same user may favor different color transformations for different web sites.* Figure 8 shows the average scores of each web site by User1 as an example. As shown in the figure, User1 prefers *Arbitrary* for CNN and Google, *Green* algorithm for Facebook, and *Dark* for Weather and ESPN. Similar results are found for other users. This, again, motivates Chameleon's design to give end user options in selecting the color transformation algorithm and selecting it per web site.

Finally, our prior work [9, 10] show that users may want different tradeoffs between energy saving and perceptual constraints, depending on the available battery level. They are willing to sacrifice more aesthetics when the need for energy saving is urgent. This again motivates that choices should be given to the end user.

## 4. Design of Chameleon

We next describe the design of Chameleon as motivated by the results from the motivational studies.

### 4.1 Key Design Decisions

The results from the motivation studies lead us to make the following major design decisions for Chameleon.

*Treat GUI Objects and Images Differently:* Chameleon only applies display darkening to foreground images in order to preserve fidelity. It applies color transformation only to GUI objects, background images, and possibly logo images depending on the user choice, according to findings reported in 3.2.2. A side benefit of only applying darkening to foreground images is that it works very well with incrementally rendered large photos.

*Keep Color Consistency per Web site:* Web pages from many web sites often employ the same color scheme to maintain a consistent user experience. Chameleon keeps this consistency by applying the same color transformation to all pages from a web site and does so for the top 20 web sites of the user. We opt against color consistency for multiple web sites because a user may prefer different color transformations for different web sites, as found in Section 3.3. Moreover, the top 20 web sites of a user accounts for most of her /his web usage according to 3.2.1.

*Generate Device Specific OLED Power Model:* As shown in Section 3.1, power models of different OLED displays are different. To make sure the transformed color scheme is optimized for each device, Chameleon builds an OLED power model for the device it runs using power readings from the battery interface.

*Calculate Color Maps Offline:* Chameleon finishes the compute-intensive mapping optimization offline, in the cloud in our implementation, and only perform the absolutely necessary tasks such as color contribution collection and painting in real-time.

*Give User Options:* For each web site, Chameleon allows a user to choose from linear and arbitrary transformations described in Section 2, to specify the color preference and perceptual constraint for the color transformation and to choose to transform logo images either by darkening or color transformation. Chameleon will have all color maps using all possible user options ready. As a result, the user will immediately see the effect of her selections without waiting. Our user study showed this is extremely useful for users to find out their favorite transformations.

### 4.2 Architecture

Now we provide an overview of the architecture of Chameleon. Without knowing the future, Chameleon uses the web usage by the user in the past to approximate that of the future. Therefore, Chameleon collects color information of web browsing and seeks to identify the color transformation for each color so that the average display power consumption of past web usage could be minimized. Cha-meleon then applies the color transformation to future web browsing to reduce the display power consumption.

Suppose a user has been browsing a web site for time $T$. Then the energy consumption in time $T$ is

$$E = \int_0^T \sum_{i=1}^N n_i(t) P_{pixel}(x_i)\, dt = \sum_{i=1}^N P_{pixel}(x_i) \int_0^T n_i(t)\,dt,$$

where $x_i = (R_i, G_i, B_i)^T, i = 1, \ldots, N$, are the $N$ colors supported by the browser and $n_i(t)$ is the pixel number of color $x_i$ in the display at time $t$. Notably, the integral factor $\int_0^T n_i(t)dt$ considers both the spatial and temporal contributions by a color. It naturally gives a larger weight to a web page that is viewed for a longer time. The integral factor can also "forget" past record by including a weight that diminishes for contributions from the distant past.

As shown in 3.1, the power consumption of an OLED pixel is a linear function of its linear RGB values, i.e., $P_{pixel}(x) = (a \; b \; c) \cdot x = a \cdot R + b \cdot G + c \cdot B$. Note that the constant factor in the original linear function is not included because it is independent from the color. Thus, the color-dependent energy consumption can be rewritten as

$$E = \mathbf{M} \cdot \mathbf{X}' \cdot \mathbf{D} = (a \; b \; c) \cdot \begin{pmatrix} R'_1 & \ldots & R'_N \\ G'_1 & \ldots & G'_N \\ B'_1 & \ldots & B'_N \end{pmatrix} \cdot \begin{pmatrix} \int_0^T n_1(t)dt \\ \vdots \\ \int_0^T n_N(t)dt \end{pmatrix}.$$

$\mathbf{M}$ is the OLED power model; $\mathbf{X}'$ is a matrix called the *color map* with the $i$th column being the transformed color for $x_i$; $\mathbf{D}$ is the color contribution vector for the web site with each entry corresponding to the contribution from color $x_i$, $\int_0^T n_i(t)dt$.

To minimize the energy consumption, $E$, Chameleon must construct the power model, $\mathbf{M}$, gather data to derive $\mathbf{D}$, calculate $\mathbf{X}'$ with user-supplied options, and apply it to change the colors of future web pages. Therefore, Chameleon consists of four modules that interact with a browser engine, as illustrated in Figure 9.

- A *model construction* module generates a power model, $\mathbf{M}$, of the OLED display of the mobile system, using the smart battery interface.
- A *contribution collection* module gathers a color contribution vector, $\mathbf{D}$, for each web site from the *Layout Calculation* and *Painting* stages of the browser engine in an event-driven manner.
- An offline *mapping optimization* module computes the color map $\mathbf{X}'$ based on $\mathbf{M}$ and $\mathbf{D}$. Note that Chameleon computes the color maps for all possible user options so that the user can immediately see the impact of a change in user options.
- An *execution* module applies the color map $\mathbf{X}'$ to transform colors in a web page.

Out of the four modules, only contribution collection and execution have to be executed in real-time. We next discuss each module in detail.

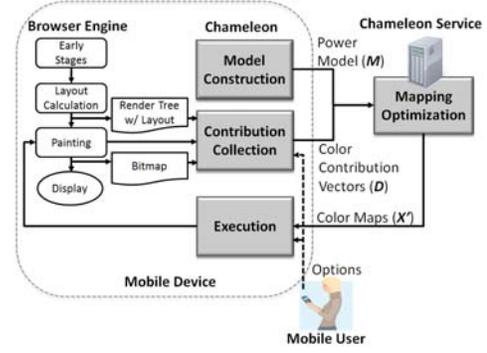

**Figure 9: Architecture of Chameleon and how it interacts with the browser engine**

### 4.3 Color Power Model Construction

Chameleon constructs the model $\mathbf{M}$ automatically without any external assistance, allowing for immediate deployment. The key idea is to employ the limited power measurement capability provided by the smart battery interface in modern mobile systems.

Chameleon's model construction module shows a series of benchmark images on the OLED display and measures the corresponding power consumption of the whole system from the battery interface. A benchmark image has all pixels showing the same color. Because all pixels have the same color, the system power consumption would be

$$P_{system} = n \times (aR + bG + cB) + P_{black},$$

where $n$ is the total number of pixels and $P_{black}$ is the system power consumption when the display is showing a black screen. When all the benchmark images have been shown and the corresponding power consumption numbers have been collected, the module applies linear regression to obtain the values of $a$, $b$, and $c$.

The model can be constructed once and calibrated over the lifetime of the device. The calibration is necessary because OLED displays are known to age and therefore exhibit different color-power properties over years. However, since the aging process is slow, the model only needs to be recalibrated using the same process a few times per year, without engaging the user.

### 4.4 Color Contribution Collection

Chameleon generates the color contribution vector, $\mathbf{D}$, of GUI objects, background images, and possibly logo images (depending on the user choice). Recall that an element of $\mathbf{D}$ is determined by how many pixels have the $i$th color and for how long. Therefore, whenever the display changes, contribution collection must determine how long the previous screen has remain unchanged, or *time counting*, and how many pixels in that screen are of the $i$th color, or *pixel counting*. To process a large number of colors ($>10^3$) and pixels ($>10^5$) in real-time, Chameleon employs a suite of techniques to improve the efficiency of contribution collection.

*Why must contribution collection be done in real-time*? It would be much easier to use the browser history to record the URLs of visited web pages with a timestamp and examine the pages offline. The key problem with this off-line method is that it does not capture what actually appears on the display. Because often a small portion of a web page can be shown on the display and the user must scroll the page, zoom in, and zoom out during browsing, the problem will lead to significant inaccuracy in the color contribution vector, *D*.

*4.4.1 Event-driven Time Counting*

Chameleon leverages the paint functions of the browser engine to efficiently count time in an event-driven manner. The browser engine updates the screen by calling a paint functions in the *Painting* stage. The time when a paint function returns indicates the screen is updated and would be a perfect time to start contribution collection. However, it is very common that the browser engine calls a series of paint functions to paint multiple nodes in the render tree with layout, even for one screen update as perceived by the user. If contribution collection runs for every paint function, the overhead can be prohibitively high. Therefore, Chameleon seeks to identify a series of paint functions that are called in a burst and only runs contribution collection when the last of the series of paint function returns.

Chameleon employs a simple yet effective timeout heuristics to tell if a paint function is the last in a series. That is, if there is no paint function called after a timeout period, Chameleon considers the last paint function the last of a series. The choice of the timeout value is important. If the timeout value is too small, contribution collection will be called frequently. If it is too large, many display updates will go unaccounted for.

To determine a reasonable timeout value, we performed a user study with ten mobile users to collect timing information of paint function calls. In the user study, each of the users freely browsed web for 30 minutes using an instrumented web browser on Nexus One. The instrumented browser records the time a paint function starts and completes, producing a 300-minute trace. Given a timeout value, the paint function calls in the trace can be grouped into series. We then calculated the timing statistics for the identified series. *Inter-series interval* is the time between the finish of the last paint function in a series and the finish of the last paint function in the next series. It is the time between two executions of contribution collection. *Series duration* is the time between the start of the first paint function in a series and the finish of the last paint function in the same series. It measures the time during which the screen updates will not be counted by Chameleon.

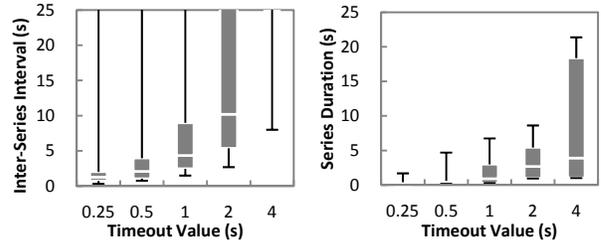

**Figure 10: Box plots for inter-series interval (left) and series duration (right) calculated using different timeout values. The bottom and top edges of a box are the $25^{th}$ and $75^{th}$ percentile; the white band in a box is the median, and the ends of the whiskers are maximum and minimum, respectively**

Figure 10 shows the box plots of inter-series interval and series duration generated using different timeout values. As shown in the figure, both inter-series interval and series duration have wide distributions. As a result, we should not only examine their averages but also their ranges. For inter-series interval, we are more interested in its lower percentile because short inter-series intervals mean frequent execution of contribution collection. When the timeout is 0.25 second, the $25^{th}$ percentile of inter-series interval is approximately 0.5s. In other words, it will be one out of four chances that the time overhead of contribution collection is over 10%. The overhead can be reduced to 2.5% with a one-second timeout. For series duration, we are more interested in its higher percentile because long series durations imply inaccuracy in the color contribution vector, *D*. When the timeout is 4s, the $75^{th}$ percentile of series duration is more than 18s. In other words, it will be one out of four chances that the screen updating for at least 18s will not be counted, which can introduce a considerable error in color contribution vector *D*. The $75^{th}$ percentile of series duration can be reduced to about 5s with a timeout of two seconds.

Therefore, we consider a reasonable timeout should be between one and two seconds and we set the timeout as one second in the reported implementations.

*4.4.2 Pixel Counting*

Once contribution collection is called, it will count the number of pixels for each color. Because the numbers of both pixels and colors can be large, we design the module as follows to improve its efficiency in both computing and storage.

Chameleon obtains pixel information, or RGB values, from the framebuffer. We note that an obvious alternative is to traverse the render tree without layout to calculate the pixel number of each color [18]. However, the render tree contains no pixel information of images, which makes it impossible to count pixels in background image. Second, the overlapping of GUI objects makes it compute-intensive to count pixels from the render tree, thanks to a possibly large number of GUI objects in a web page. For example, counting the pixel number of all the colors in the web page shown in Figure 6 (a) from the render tree with layout takes more than 200ms, compared to less than 60ms it takes to from the framebuffer.

Figure 11 shows the whole process of pixel counting. As in Step (1) of the figure, Chameleon copies the screen content from the framebuffer to the main memory in a stream and examines pixels in the main memory copy to minimize expensive framebuffer accesses. Moreover, Chameleon only reads from the framebuffer for the screen region that has been updated by the last series of paint functions. Because each paint function includes parameters that specify a rectangle region in the screen on which it draws, Chameleon calculates the superset rectangle of these rectangles and only updates the super set to the main memory copy.

Chameleon excludes foreground images from the rectangular superset. Chameleon leverages two parameters of a paint function to identify the area of the superset rectangle that belongs to foreground images. One parameter indicates what low-level library to use and this parameter tells whether the pixels to be drawn are from images or GUI objects. The other parameter is the pointer of the image node in the render tree with layout. Using this pointer, Chameleon tells if the image is foreground or background by its position in the render tree with layout: a foreground image is an element node of the render tree while a background image is a property of an element node. In case of a foreground image, Chameleon reads its size and position and skips its pixels in pixel accounting.

Chameleon employs a Hash table to store the color contribution vector, $D$, to reduce its storage requirement because the number of colors used by a web site is orders of magnitudes smaller than that of all possible colors. To reduce the index collision and make an efficient Hash table, we choose to use a special key for the Hash table, i.e., each of the RGB components is reversed in bit and then interleaved together. The rationale of such a choice is that, compared to lower bits, the higher bits of the RGB components are more effective to differentiate colors and should be used earlier by the Hash table. The Hash table implementation only requires about 10KB for a web site because the number of colors used by a web site is no more than a few thousands. When updating the Hash table, Chameleon can discount the existing record, instead of simply aggregating the new pixel counts into it, in order to "forget" web pages browsed in the distant history.

As in Step (2) of Figure 11, Chameleon counts pixels in the main memory copy by 10 pixels by 10 pixels (10×10) blocks and employs a small Hash table with special collision resolution to store the results for a block. The key of the small Hash table is the higher two bits of RGB components of each color; the table resolves a collision by merging the existing entry into the per-website Hash table, as in Step (3) of Figure 11. After all the blocks have been examined, Chameleon then merges the small Hash table into the per-web site Hash table, as in Step (4) of Figure 11. This design is motivated by the spatial locality of colors in web pages: the average color numbers in a 10×10 block is only 5 according to the LiveLab traces. By counting pixels block by block and using the small Hash table of a fixed size to store the block results, Chameleon reduces the number of writings into the per–website Hash table.

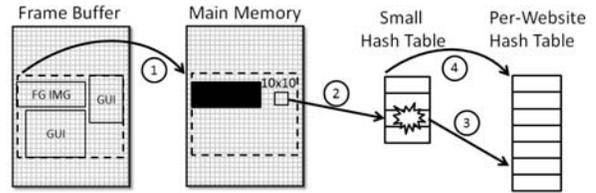

**Figure 11: How Chameleon counts pixels**

Finally, Chameleon can further leverage the spatial locality of color to reduce the number of pixels to examine by sampling. That is, instead of counting every pixel, Chameleon can only count one pixel out of a block (e.g., 10×10). This pixel can be sampled from a fixed position or simply by random. This sampling technique makes a profitable tradeoff between the accuracy of $D$ and the efficiency. Our early work, reported in [18], suggests that a sampling rate as high as one out of 10×10 pixels can still yield very high accuracy (95%) in display power estimation, thanks to the spatial locality of color in web pages.

As will show in Section 6, the time overhead of the execution of contribution collection is at most 66ms and 17ms without and with sampling.

### 4.5 Color Mapping Optimization

Chameleon treats foreground images and GUI objects separately. It only darkens foreground images but transforms the color schemes of GUI objects, background images, and possibly logo images (depending on the user choice), all under a perceptual constraint.

Given $M$, the power model, $D$, the color contribution vector, Chameleon calculates the optimal color map, $X'$, offline. Chameleon will compute the color maps for all possible combinations of user options, including all possible perceptual constraints. With all color maps at hand, Chameleon allows the user to change the options and see the resulting color scheme immediately. There are 20 color maps for each web site in our implementation.

### 4.6 Color Transformation Execution

Given a color map $X'$, Chameleon replace each original color with its corresponding transformed color for GUI objects, background images, and possibly logo images (depending on the user choice). Instead of transforming pixel by pixel, Chameleon modifies the parameters of paint functions to transform the entire region updated by the function at once. In the Painting stage, when a geometry primitive is painted with a color $(R, G, B)$, Chameleon uses $X'$ to find the transformed color $(R', G', B')$ and passes $(R', G', B')$ to the low-level graphics library to continue the painting process.

For foreground photo images, the transformation only involves darkening, or reducing brightness, i.e., each pixel $(R, G, B)$ in the image becomes $(\lambda R, \lambda G, \lambda B), \lambda \in [0,1]$, in the transformed image. This darkening operation is integrated into image decompression process such that the RGB components of each pixel are multiplied by $\lambda$ right after they are calculated by the image library.

# 5. Implementation of Chameleon

We next report our implementations of the Chameleon design described above. The first choice we faced was whether to implement it an add-on to existing web browsers or to directly modify the source code of a browser.

Many web browsers support add-ons to enhance functionality as either a plug-in or an extension. Plug-ins are provided as libraries by third party software companies to handle new types of contents, such as Flash and Quicktime movies, and unable to impact on how a web browser renders a web page. Therefore, Chameleon cannot be implemented as a plug-in. Extensions are designed by users using XML and JavaScripts to add or change browser features they prefer. An extension can use a JavaScript to change the color of any GUI object. We opt not to implement Chameleon as an extension for three reasons. First, such an extension is unable to generate a color contribution vector as described in Section 4.4 because JavaScripts have no access to the render tree with layout to obtain layout information of GUI objects and images. Second, an extension cannot transform color of background images because JavaScripts do not affect how images are decompressed. Third, an extension-based implementation will be prohibitively expensive because the extension needs to traverse the whole render tree to execute color transformation. For example, to perform a color inversion on CNN mobile home page using such a script costs more than 500ms while Chameleon only costs only 2ms.

We choose to implement Chameleon by modifying the source code of a web browser. We have implemented Chameleon on Fennec, the mobile version of Mozilla Firefox, and the Android WebKit. WebKit is an open-source web engine which serves many modern web browsers such as Safari and Chrome. One can change the painting stage of WebKit to realize Chameleon. Because an Android system includes WebKit as part of the kernel, a modification of the WebKit codebase requires rebuilding the whole Android system. Fennec is an open source web browser project that is ported on both Android and Maemo operating systems. Unlike the Android Web browser, Fennec is a stand-alone application and can be installed on almost any mobile platforms with Android/Maemo systems without rebuilding the system. The downside of Fennec is that it requires storage of over 45MB.

Due to page limits, we only report the details of Fennec implementation because it does not require rebuilding Android from source. The Android WebKit implementation, however, is similar. Our modification to the Fennec includes changing 15 files and adding/changing 387 lines of code. Chameleon/Fennec runs on any Android/Maemo based mobile platform. We next use Nexus One as an example target device to present the implementation details.

## 5.1 Automatic Power Model Construction

Battery interface of Nexus One updates a power reading in every 56 seconds, which is computed by averaging the last 14 samples (1 sample per 4 seconds). Therefore, it is important to make sure Nexus One is operating with a with stable power behavior during the model building process. We enforce several methods to achieve a stable power behavior, including turning off all the wireless links by using airplane mode and shutting down all the running $3^{rd}$ party applications except Chameleon. More importantly, the calibration process itself, updating framebuffer using different colors, should have a stable power behavior. Chameleon employs OpenGL to update the framebuffer in order to minimize energy consumption in non-display components, e.g. storage access when a gallery application is used to present images. Because the linearity of color power model, it is unnecessary to go through all the colors. We only calibrate sixteen levels for each of the red, green, and blue. Each level costs 56s and the whole model building process takes approximately one hour. A user should start such model building process from the menu before using Chameleon.

## 5.2 Color Contribution Collection

Contribution collection is implemented as a function and is called by Chameleon after a timeout since a paint function returns as described in Section 4.4. When contribution collection is called, Chameleon reads the updated screen region from the framebuffer to the main memory in a stream using OpenGL function GLReadPixels. The time taken by this copying process depends on the region size and is at most 16ms in Nexus One. As described in Section 4.4, the color contribution vector for a web site, $D$, is implemented as a Hash table. The Hash table for each web site is saved as a binary data file. When Chameleon opens a web site, it loads the corresponding binary data file and updates it accordingly. A typical size of such a binary file is ~6KB such that loading/saving the Hash table will not introduce much overhead.

We note that color contribution collection does not need to be always on. As we will see in Section 6.3, two weeks of color contribution collections will lead to close to optimal power reduction for at least the next three months for an average user.

### 5.3 Color Mapping Optimization

Since mapping optimization runs offline and is independent from Chameleon, we choose to implement it as a service running in an Internet server. This service takes two inputs, i.e., a triplet of three float numbers representing the OLED display power model of a Chameleon user's mobile device and a binary data file consisting of 20 color contribution vectors of the user's top 20 most visited web sites. The user can initiate the mapping optimization service from Chameleon's menu. Chameleon will upload the two inputs to the Chameleon server and the service will calculate optimized color maps using all possible algorithms and options. For each web site, the service generates color maps using the four algorithms described in Section 3.3, five parameters settings for each algorithm. As a result, there are 20 color maps for each web site. Then Chameleon automatically downloads this output data file to the mobile device. Finally, a color map will be selected for each website based on the user's choice. When the user changes his choice, a new color map will be selected and applied immediately without needing the server.

The mapping optimization service is implemented in two parts: a front-end interface implemented by PHP to handle requests from Chameleon and a back-end computation engine. The back-end engine is implemented in C++ and employs the GNU Scientific Library for optimization.

### 5.4 Color Transformation Execution

Chameleon implements color transformation of GUI objects by modifying the color interface of Fennec. When a paint function draws a geometry primitive, i.e., a point, a line or a rectangle, it uses a HEX string or a color name as a parameter to specify the color of the primitive. The color interface translates such a HEX string or a color name to RGB values that will be further used by the low level graphics library. We add code right after where the RGB values are calculated. The added code uses the RGB triplet as address to find a transformed color from the color map. Since we have modified the color interface to return the transformed RGB values instead of original ones, the paint function will draw the geometry primitive using the transformed color. Therefore, the overhead induced by color transformation execution only involves the time to load a RGB triplet from the color map. As will show in Section 6.4, the total time overhead of execution in opening a web page is less than 5ms on average, which is negligible than the total time to open a web page (2-8s according to [12]).

Chameleon implements color transformation for images by modifying the image library of each individual image format because each image library employs a unique decompression procedure. In particular, code is added to modify the RGB values of image pixels after the decompression finishes. The new RGB values are determined based on if the image should be darkened or color-transformed.

## 6. EVALUATION

We evaluate through measurement, trace-based emulation, and multiple field trials. The evaluation shows that Chameleon is able to reduce OLED smartphone power consumption by 41% during web browsing.

### 6.1 Experimental Setup

Google Nexus One is used in the measurement and trace-based emulation. We run a script that automatically feeds Chameleon with web pages specified by a URL list extracted from the LiveLab traces. We measure time overhead of Chameleon by inserting time-stamps in the Chameleon source code and counting the latencies contributed by the contribution collection and execution modules. We obtain the power consumption of Nexus One by measuring the battery current through a sensing resistor with a 100Hz DAQ from Measurement Computing.

### 6.2 Power Model Accuracy

To examine the accuracy of OLED power model generated by Chameleon, we randomly select 100 web pages from the LiveLab traces and compare the power number estimated by the power model against DAQ measurement. The results show that 95% of the error is within ±10% and the average absolute error is 3%.

### 6.3 Power Reduction

As mentioned in the beginning of Section 4, Chameleon transforms future web pages based on color transformations optimized with past web usage. This raises two questions: (*i*) how long does a user need to train Chameleon; and (*ii*) how well does the past predicts the future? By "train," we mean to run the contribution collection module of Chameleon to gather the color contribution vector, ***D***.

We leverage the LiveLab traces, the display power model of Nexus One, the *Arbitrary* color transformation used in the study reported in Section 3.3, to answer these two questions. We first train and test Chameleon week by week, using the same week's trace to train and test. The resulting power reduction is not realistic but serves as a theoretical upper bound. The weekly average for all 25 users is shown in Figure 12 as *Optimal*. Then we train Chameleon using first one to four weeks of traces and test it using the rest of the traces, respectively. Again the weekly averages for all users are shown in Figure 12. Because the color transformation in Chameleon is per-web site, the weekly average is calculated over the top 20 web sites for all users. We note this emulation is an approximation only because the LiveLab traces did not actually capture what appeared on the display, as discussed in Section 4.4.

As shown in Figure 12, power reduction with two to four weeks training is close to each other but that with one week training is the obviously lowest. This indicates that *two weeks of training should be enough for Chameleon on average*. The average display power reduction is 64% and such reduction remains close to the optimal (~70%) for the rest of traces (or at least 10 weeks), as shown in Figure 12. So *the past doest predict the future well*. We also note that there is a not-so-obvious trend of declination in the power reduction with two to four weeks training only over the long term. This suggests that Chameleon only needs to train (or run the contribution collection module) for two weeks a few times a year to maintain its effectiveness in power reduction.

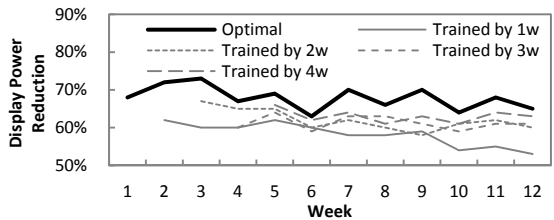

Figure 12: Display power reduction by training Chameleon with different weeks of data from the LiveLab traces

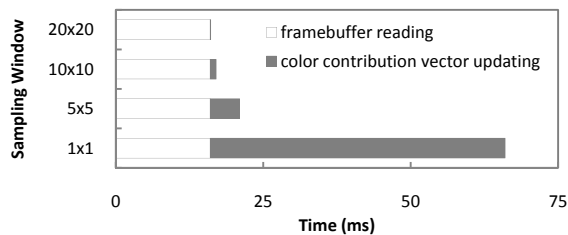

Figure 13: Worst case time overhead of Chameleon

We also measured the system power consumption by Nexus One when browsing web pages in the LiveLab traces. The average system power consumption is 1.3W and 2.2W with Fennec and Chameleon, respectively. This indicates over 41% total system power reduction through Chameleon.

### 6.4 Overhead

We next examine the overhead in time and power introduced by Chameleon. For time overhead, we are only interested in contribution collection and execution because only they run in real-time during web browsing. Our measurement shows that execution takes less than 5ms on average in opening a web page, which is negligible compared to the total time to open a web page in a mobile browser (2-8s according to [12]). The time overhead from contribution collection consists of two parts: reading the framebuffer and updating the color contribution vector, $D$. Figure 13 shows the worst case overhead of both parts using different sampling sizes. As shown in the figure, framebuffer reading costs less than 16ms in the worst case, i.e., reading the whole screen. The average reading time is ~12ms because only part of the framebuffer is read each time. The figure also shows that cost for updating the color contribution vector is ~50ms without sampling and <1ms using a sampling window of 10×10. Therefore, the total time overhead of color contribution vector updating is at most 66ms without sampling and <17ms using a sampling window of 10×10.

Such overhead will be barely noticeable given the fact that it takes 2-8s to open a web page on a mobile browser [12], as confirmed by our own experience and the field trial to be reported in Section 6.5. Moreover, the overhead of Chameleon is likely to be reduced as smartphone hardware becomes better while the web page opening time will remain large because it is determined by network condition other than smartphone hardware [12]. Finally, as shown in Section 6.3, Chameleon only needs to run contribution collection and incur this overhead for two weeks a few times a year to be effective.

We also measured the power overhead of Chameleon by going through the LiveLab URL list twice, one with Fennec and the other with Chameleon. To eliminate the power difference introduced by the display, a color map that does not change the color at all is used in Chameleon. The results show that the system power difference between the two trials is <5%, negligible compared to the 41% reduction from Chameleon.

### 6.5 Field Trials

Two field trials have been performed with Chameleon/Fennec during two points of the development. The first trial was with two Nexus One users and an early version of Chameleon. The two users used Chameleon for one week and provided valuable feedbacks. For example, the issues with background images and logo images discussed in Section 3.2 were reported by the two participants whose different preferences also motivated us to give the end user an option in how to treat logo images.

The second trial was with five users of Nexus One and Samsung Galaxy S and the reported implementation of Chameleon. The trial lasts seven days for each user. After the trial, we asked each user to use two Nexus One smartphones, one with the original Fennec and the other with Chameleon. They used the smartphones in the lab at a random order to access the same web site of their favorite in order to assess if there are noticeable latency introduced by Chameleon. No user noticed any slowdown in web browsing in Chameleon compared with using the original Fennec. From our post-trial interview of the users, we found that the ability to see the effect of changing a user option is very important, as Chameleon currently supports. Finally, all the users are satisfied with the aesthetics of the transformed color scheme of her/his choice. These early results suggest Chameleon is effective in reducing power reduction while keeping the user happy.

## 7. Discussions

Chameleon does not consider video right now because videos are not yet widely supported by mobile browsers. For example, Safari on iPhone has no support of flash, a typical video container in web pages. However, supporting video is also easy for Chameleon as videos can be considered in a way similar to foreground images. Chameleon can easily apply darkening to videos by leveraging the *opacity level* defined in CSS standard. Opacity level is an property of each object and in the range of [0,1]; it is used to handle the overlapping of two objects. To darken a video by a scale factor $\lambda \in (0,1)$, Chameleon can overlay a black image with the opacity level of $(1-\lambda)$ of the same size on the top of the video.

## 8. Conclusions

In this work, we report the design and realization of Chameleon, a color-adaptive mobile web browser to reduce the energy consumption by OLED mobile systems. Chameleon is able to reduce the system power consumption of OLED smartphones by over 41% for web browsing, without introducing any user noticeable delay.

We found studying the users and the web usage by users in the field instrumental to the design and success of Chameleon. Much of Chameleon's optimization techniques were directly motivated by findings from these studies.

Emulation using the LiveLab traces showed that two weeks of web usage is enough for Chameleon to derive a color transformation that performs close to the optimal for at least ten weeks. This suggests that only the execution module of Chameleon needs to be performed all the time and the color contribution collection module only need to be performed for two weeks once or twice a year.

Chameleon also represents a good example of how the "cloud" can be leveraged for the usability and efficiency of mobile devices. By cleverly offloading compute-intensive mapping optimization to the cloud and obtaining all possible color maps, Chameleon allows the user to see the visual impact of different user options without delay.